\documentclass{article}

\usepackage[numbers,sort&compress]{natbib}

\usepackage{amsbsy} 
\usepackage{amssymb}
\usepackage{graphicx}
\graphicspath{{./}{./Figures/}} 
\usepackage{subfig}
\usepackage{multirow} 
\usepackage{multicol}
\usepackage{booktabs}	
\usepackage{setspace}
\usepackage{xspace}
\usepackage{comment}
\usepackage[margin=2cm]{geometry}
\usepackage{lineno}		

\date{}

\usepackage[plainpages=false,
colorlinks,
debug,
linkcolor=red,
citecolor=red, 
filecolor=red,
breaklinks=false]{hyperref}

\usepackage{glossaries}
\glsdisablehyper 
\setacronymstyle{long-short}
\newacronym{bam}{BAM}{Bundesanstalt f\"{u}r Materialforschung und -pr\"{u}fung, Berlin, Germany.}
\newacronym{bf}{BF}{bright-field}
\newacronym{df}{DF}{dark-field}
\newacronym{ecap}{ECAP}{equal channel angular pressing}
\newacronym{eds}{EDS}{energy dispersive X-ray spectropscopy}
\newacronym{fib}{FIB}{focused ion beam}
\newacronym{grf}{GRF}{growth restriction factor}
\newacronym{haadf}{HAADF}{high angle annular dark-field}
\newacronym{hpt}{HPT}{high-pressure torsion}
\newacronym{hrtem}{HRTEM}{high resolution transmission electron microscopy}
\newacronym{hv}{HV}{Vickers hardness}
\newacronym{pals}{PALS}{Positron annihilation lifetime spectroscopy}
\newacronym{pips}{PIPs}{precision ion polishing system}
\newacronym{rlm}{RLM}{reflected light microscopy}
\newacronym{rms}{RMS}{root mean square}
\newacronym{sadp}{SADP}{selected area diffraction pattern}
\newacronym{saxs}{SAXS}{small angle X-ray scattering}
\newacronym{sem}{SEM}{scanning electron microscopy}
\newacronym{spd}{SPD}{severe plastic deformation}
\newacronym{stem}{STEM}{scanning transmission electron microscopy}
\newacronym{tem}{TEM}{transmission electron microscopy}
\newacronym{ufg}{UFG}{ultrafine-grained}
\newacronym{uts}{UTS}{ultimate tensile strength}
\newacronym{xrd}{XRD}{X-ray diffraction}

\usepackage{authblk}

\title{Dynamic precipitation during high-pressure torsion of a magnesium-manganese alloy}
\author[1]{J. M. Rosalie\thanks{julian.rosalie@bam.de}}
\author[2]{A. Hohenwarter}
\affil[1]{Bundesanstalt f\"ur Materialforschung und -pr\"ufung,
	Unter den Eichen~87,
	12205,
	Berlin,
	Germany}
\affil[2]{Department of Materials Science, Technical University of Leoben,	Jahnstra{\ss}e 12,	Leoben, 8700, Austria}

\begin{document}
	
	\maketitle

	\begin{abstract} 
		An ultrafine grained magnesium alloy has been produced through room temperature \gls{hpt} of solutionised Mg-1.35 wt\%Mn. Dynamic precipitation of nanometer-scale Mn particles occurred during deformation. These particles populated the grain boundaries, acting as pinning sites which allowed the alloy to develop a  grain size of 140\,nm after 0.5 rotations. Further \gls{hpt} deformation resulted in a gradual increase in grain size with no increase in precipitate size. Despite the extensive deformation applied, the alloy did not develop a bimodal grain structure and  retained a grain size of 230\,nm after 10 complete rotations, demonstrating the  stability and effectiveness of these pinning particles.
	\end{abstract}


\section{Introduction}	

\glsreset{hpt}

\Gls{spd} has been widely studied as a means to refine the grain size of metallic materials down to tens or hundreds of nanometers. This has been shown to be beneficial not only in terms of increasing strength via reducing the mean-free path for dislocations, but also to potentially developing new phases, enhancing diffusion and modifying surface behaviour\cite{HAHN2015101, Edalati2022,ValievEstrin2016}.

Numerous researchers have applied \gls{spd} to high- or commercial-purity magnesium \cite{Edalati2011,Sulkowski2020,QiaoZhao2014} and several of its alloys\cite{XiaWang2005,Harai2008,XuShirooyeh2013,Silva2019,Yu2015}. The minimum grain size of pure magnesium obtainable through room temperature \gls{spd} is limited to approximately 0.6--1$\mu$m due the high homologous temperature \cite{Edalati2011,QiaoZhao2014}, and subsequent dynamic recrystallisation and grain growth during deformation. Finer grain sizes of 0.1--0.2\,$\mu$m  have been achieved with common structural  grades of magnesium, such as the AZ \cite{XuShirooyeh2013,XiaWang2005,Harai2008} and ZK series\cite{Silva2019} and these have potential in transport and similar applications requiring high strength with low weight.

In addition to its use as a lightweight structural material, magnesium is also of value in the biomedical field for use as a biodegradable implant material\cite{GuZheng2010,ValievEstrin2016,YangYoon2015,GutierrezPua2023, FigueiredoLangdon2019}. Critical considerations  are tailoring the balance between strength and corrosion resistance\cite{WANG2011579} and optimising the \textit{in vivo} dissolution rate. This is complicated by the fact that few of the common alloying elements used in magnesium alloys are well-tolerated within the body. 
\Gls{spd} of magnesium alloys has been found to show promise as a means of augmenting the mechanical properties without detriment to the biocompatibility or corrosion resistance  \cite{Medeiros2023}. Magnesium alloys also have potential as hydrogen storage materials\cite{Edalati2011,QiaoZhao2014,JORGE20138306,FigueiredoLangdon2019}, where a fine grain size is critical for hydrogen absorption \cite{Edalati2011} where the challenge is increasing the absorption/desorption rate for hydrogen, while retaining thermal stability. The common feature of both these applications is a need for improved control of the grain size.

One approach to limiting grain growth and achieving finer grain sizes during \gls{spd} is the introduction of pinning particles.  Magnesium-manganese alloys (often described by their ASTM designation of ``M$x$´´ for $x$ wt.\%Mn content \cite{PrasadBhingole2017})  are a promising system from this perspective, as rod-like precipitates of $\alpha$-Mn\cite{StratfordBeckley1972}, nucleated on dislocations\cite{Skjerpe1983},  have been shown to limit recrystallisation in coarse-grained materials, and were themselves resistant to dissolution during rolling and annealing\cite{Robson2011}. Similarly, Mn particles precipitated during homogenisation have been shown to retard grain growth during extrusion of Mg-Gd alloys \cite{FangYi2009}. Mg-Mn alloys  are also employed commercially as a  creep-resistant materials in the nuclear industry \cite{CelikinKaya2012,Borkar2012,CelikinKaya2012a}, due to the ability of Mn particles to dynamically precipitate on dislocations during creep deformation.

\Gls{spd} of Mg-Mn alloys has been studied using by via extrusion\cite{Yu2015,SomekawaBasha2018,Borkar2012} and  \gls{ecap}\cite{SvecDuchon2012}. Extrusion of an M1 alloy produced a strongly textured microstructure with a grain size of $\sim$3.1\,$\mu$m and a combination of high strength, anisotropy and room temperature ductility \cite{Yu2015}. A study using \gls{ecap} \cite{SvecDuchon2012} reporting grain sizes of 0.5 and 0.6 microns in Mg-0.2 wt\% and 1.5 wt\% Mn, however the \gls{uts} values of 140\,MPa and 110\,MPa were relatively low; a factor attributed to the strong texturing common during \gls{ecap} deformation of Mg. In these studies the alloy had been slowly cooled to room temperature prior to deformation, and given the low solid-solubility at room, temperature, the majority of the Mn would be partitioned into coarse, widely-separated intermetallic particles. 

Whereas previous studies have generally considered the effect of pre-existing $\alpha$-Mn particles on grain refinement  of cast\cite{YuTang2014}, pre-heated \cite{YuTang2018} or furnace-cooled\cite{SvecDuchon2012} alloys, the present study investigated \gls{hpt} deformation of a solutionised Mg-Mn alloy. Dynamic precipitation has been reported  during \gls{hpt} of solutionised Al-Cu alloys \cite{Hohenwarter2014, Nasedkina2017}, but to the best of the authors knowledge has not been considered in solutionised Mg-Mn. The present study therefore aims to capitalise on the high defect density during \gls{hpt} combined with the ability of Mn to precipitate dynamically on such defects to develop a particle-rich microstructure capable of restricting grain growth.  This was driven by the fact that  manganese is well-tolerated in the human body, and a sufficiently robust, adequately corrosion resistant alloy would have potential as a resorbable biomaterial. This study involved a detailed investigation of grain size and precipitation via electron microscopy.

\section{Experimental details}

The alloy used was a commercial grade M1 provided by Magnesium Elektron, the compositional analysis of which is shown in Table~\ref{tab-composition}. Specimens of the as-received material were prepared using standard metallographic techniques and examined using \gls{rlm}  and \gls{sem}.  Grain sizes were measured via \gls{rlm} using a linear-intercept method, on samples etched with a mild nitric/acetic acid etchant\cite{Maltais2004}.
\gls{sem} samples were embedded in cold-mount resin with a conductive filler and given a final polish using a Vibromet 
vibratory polisher immediately prior to examination in a FEI Quanta 3D \gls{sem}/\gls{fib}.

\begin{table}[hbtp]
	\setlength{\tabcolsep}{2pt}
	\begin{center}
		\caption{Chemical composition of the M1 alloy. (RE=Rare earths) \label{tab-composition}}
		\begin{tabular}{*9{r}}
			\toprule
			Elem. 	& Mn 		&	Cu		&	Si	& Ni 	& Ca		& RE & Other &Mg \\
			wt.\% 	&1.35		& 	0.000 	&     0.002 &0.000 & 0.00 & 0.00 & $<$0.01 & bal.\\
			\bottomrule
		\end{tabular}
	\end{center}
\end{table}

\begin{figure}
	\centering
	\includegraphics[width=0.48\textwidth]{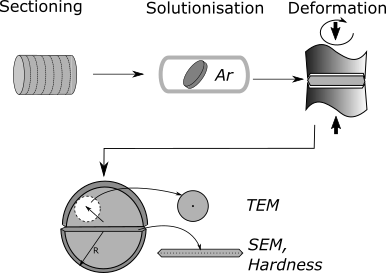}
	\caption{A schematic of the process of sample preparation and analysis. \label{fig-processing}	}	
\end{figure}

The process of \gls{hpt} sample preparation and analysis is illustrated schematically in Figure~\ref{fig-processing}. As-received material was machined into 8\,mm diameter cylinders, and sectioned into 1\,mm thick discs. These specimens were solutionised in sealed Argon-filled glass ampoules for for 8 hours at 900\,K  and water-quenched. Deformation was applied using a Schenk high pressure torsion apparatus operating at a pressure of 7.5\,GPa. Discs were held within tool steel anvils, with an anvil gap of diameter 8\,mm and cavity depth  0.15\,mm. Processing was performed at room temperature, with a rotational velocity of 0.4\,revolutions per minute. After deformation the disc thickness was  0.50$\pm$0.01\,mm. 

\gls{hpt} deformed discs were sectioned using a diamond-wire saw and embedded in cold-mount resin. Vickers microindentation hardness testing was conducted using a QATM Qness 60 A/A+ EVO hardness tester with a 50\,g load. Indents were made in the tangential direction along the mid-plane of the disc.

Specimens were then extracted from the resin, mechanically thinned and dimple ground. These were then thinned to perforation using a Gatan \gls{pips} II, using liquid nitrogen cooling. The central, electron-transparent of each disc corresponds to a radial distance of approximately 2.0\,mm during deformation. 

\Gls{tem} and \gls{hrtem} observations were carried out using a JEOL 2200FS instrument operating at 200\,kV.
\Gls{stem} measurements were conducted using a Thermo-Fischer Talos F200S microscope, also at 200\,kV. Grain sizes were determined using the line-intercept method applied to \gls{bf} \gls{stem} images.

\section{Results}

\subsection{Mechanical properties} 

\begin{figure*}[hbtp]
	\begin{center}
		\includegraphics[width=12cm]{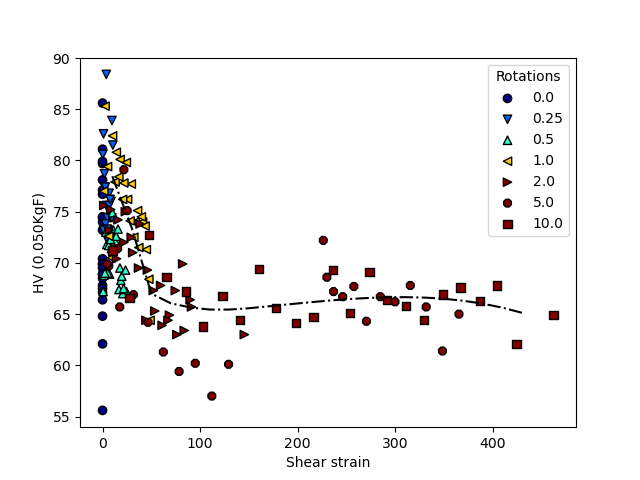}
		\caption{Vickers indentation hardness of \gls{hpt}-deformed Mg-Mn versus applied strain. The line is included as a guide for the eye, only.\label{fig-hardness}}
	\end{center}
\end{figure*}

Compressive deformation resulted in an increased, but highly variable hardness value of 72$\pm$7\,\gls{hv}, compared to a value of 46$\pm$\,3\,\gls{hv} in the as-received condition. This variability was ascribed to the inhomogeneous strain distribution after compressing the disc into the anvil cavity.
The alloy underwent further increases in hardness upon \gls{hpt} deformation,  reaching 88$\pm4$\,HV at 0.25 rotations. 
However, subsequent shear deformation resulted in an unexpected decrease in hardness and at $\ge$2 rotations the Vickers hardness plateaued at 66$\pm$3 HV. 
A composite plot showing the hardness for all samples  is presented in Figure~\ref{fig-hardness}. The plot shows the microindentation hardness versus shear strain.
$ \gamma = \frac{2\pi r N }{t} $
where the strain, $\gamma$ for a given number of rotations, $N$ is determined from the radial distance $r$ and the disc thickness, $t$.  The strain for the compressed sample was calculated from the logarithmic strain for the height reduction of the discs from $\sim$0.85\,mm to 0.50$\pm$0.01\,mm, giving a value of 0.5.

\subsection{Microstructure}

\subsubsection{As-received condition}

\begin{figure*}[htbp]
	\centering
	\hfill
	\subfloat[\label{fig:mgmn-tem-summary-sem}]{\includegraphics[width=0.32\textwidth]{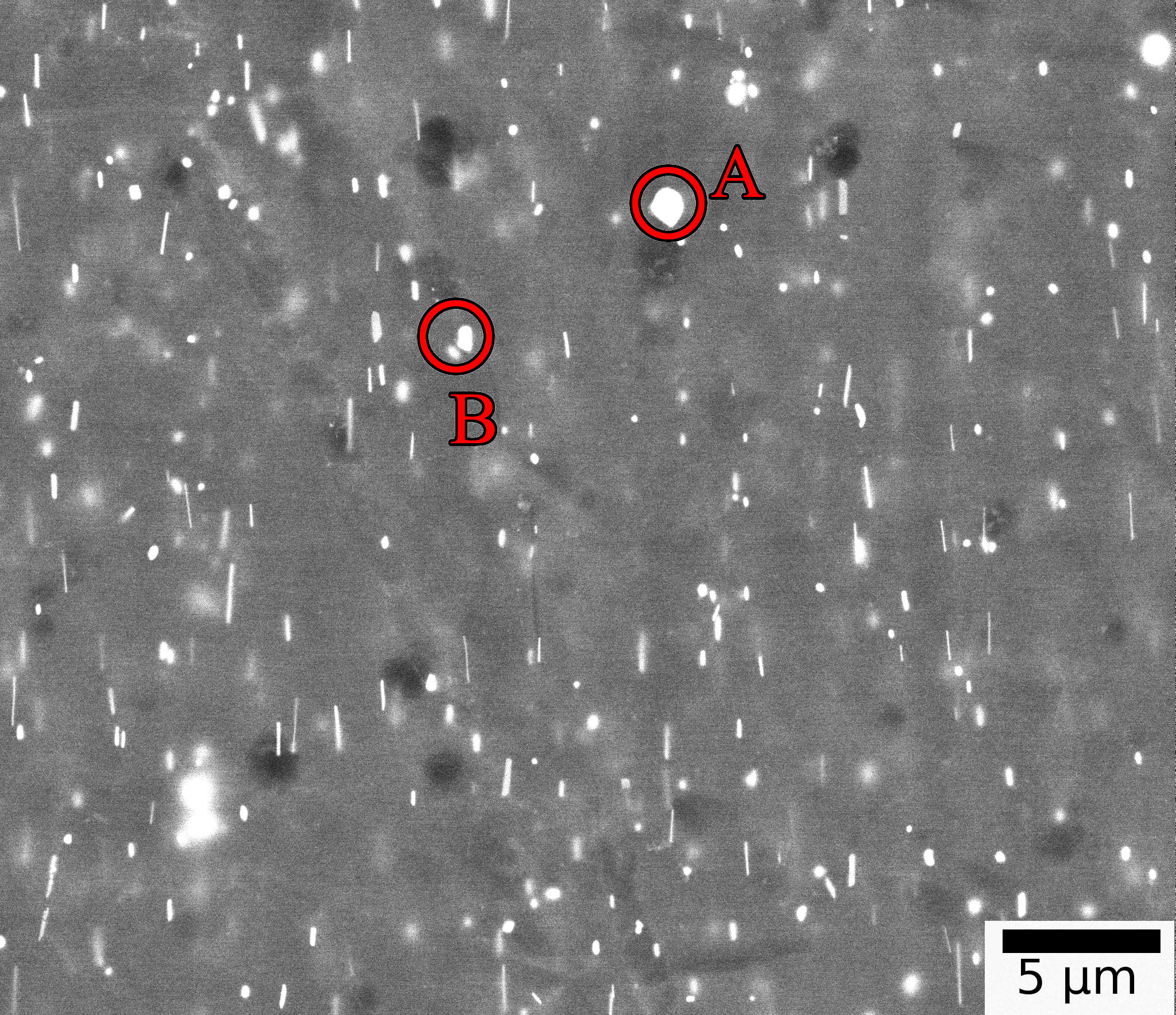}}
	\hfill
	\begin{minipage}[b]{0.32\textwidth}{
				\subfloat[\gls{eds} of particle A \label{fig:mgmn-tem-summary-sem-eds-1}]{
				\includegraphics[width=\textwidth]{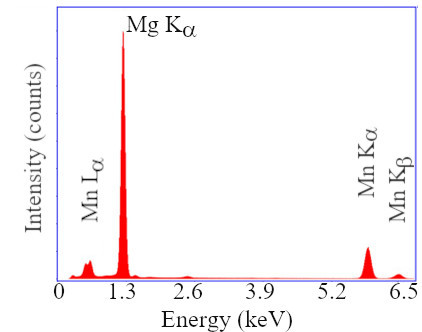}}
			
			\subfloat[\gls{eds} of matrix\label{fig:mgmn-tem-summary-sem-eds-2}]{
				\includegraphics[width=\textwidth]{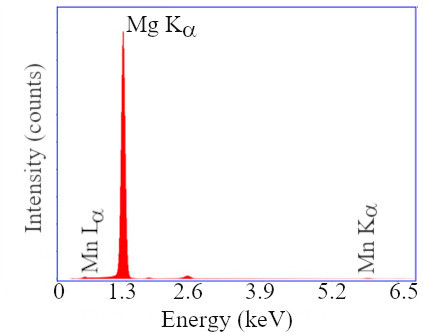}
		}	}
	\end{minipage}
	\hfill\
	\subfloat[\label{fig:mgmn-tem-summary-ar}]{\includegraphics[width=0.32\linewidth]{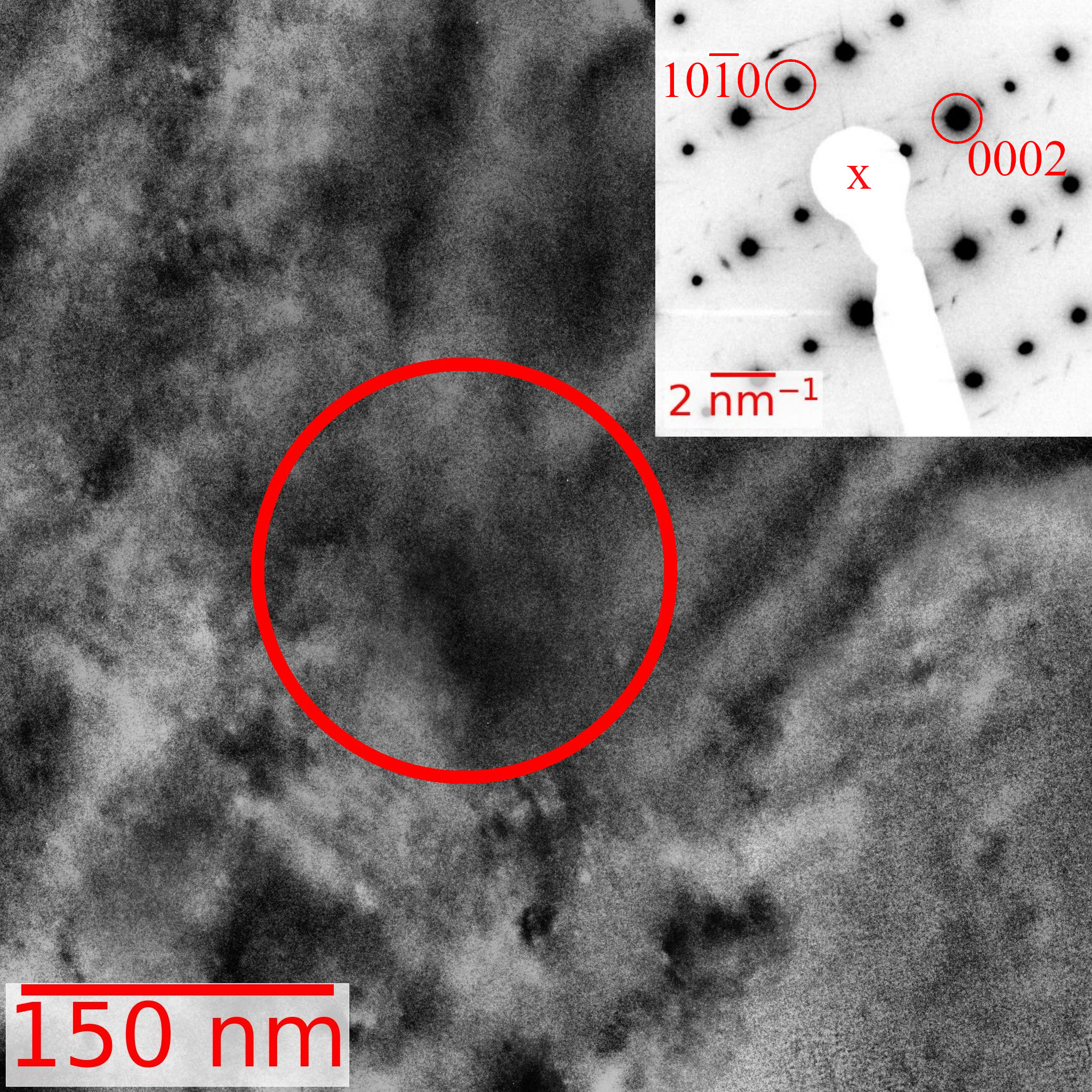}}
	\caption{The microstructure of the as-received alloy, prior to solution-treatment and deformation.
		(a) \gls{sem} back-scattered electron image showing coarse rod-like, and occasional spheroidal,  particles of $\alpha$-Mn. 
		\gls{eds} spectra from (b) the particle marked ``A" in the \gls{sem} image,  and (c) the matrix, showing a higher concentration of Mn at the particle, and the absence of additional elements.
		(d) \Gls{bf} \gls{tem} image. The inscribed circle shows the aperture position for the inset 
		\gls{sadp},  which indicates this grain is oriented along the $11\overline{2}0$ direction, and which shows weak reflections assigned to MgO but no evidence for the existence of Mn-containing phases.
		\label{fig:mgmn-tem-summary}
	}  
\end{figure*}

Coarse rod-shaped and spheroidal Mn particles can be clearly seen in backscattered \gls{sem} images of the as-received condition (Fig~\ref{fig:mgmn-tem-summary-sem}), as expected from experimental and theoretical studies which indicate that Mn has minimal solubility in Mg at room temperature \cite{GroebnerMirkovic2005, Kang2007}. \Gls{eds} showed a higher intensity of  Mn K$_\alpha$ X-rays at such particles (\ref{fig:mgmn-tem-summary-sem-eds-1}), than in the adjacent matrix (\ref{fig:mgmn-tem-summary-sem-eds-2}). \gls{bf}-\gls{tem} images (Fig.~\ref{fig:mgmn-tem-summary-ar}) did not detect finer Mn particles than those visible in the \gls{sem} images. \Glspl{sadp} (inset) show  reflections due to MgO hkl=200 and 220 reflections (presumably indicating minor surface oxidation) but not Mn in regions where coarse particles were absent, showing that Mn is concentrated in the form of these coarse particles, and that there is no indication of nanoparticles of Mn in the as-received state, i.e. prior to solution treatment. A grain size of 26$\pm$1$\,\mu$m was measured via \gls{rlm}.

\subsubsection{As-compressed}

\begin{figure*}
	\centering
	\hfill	

	\subfloat[\label{fig:20230531-mgmn-0rotn-a3-tem-bulk}]{\includegraphics[width=0.32\linewidth]{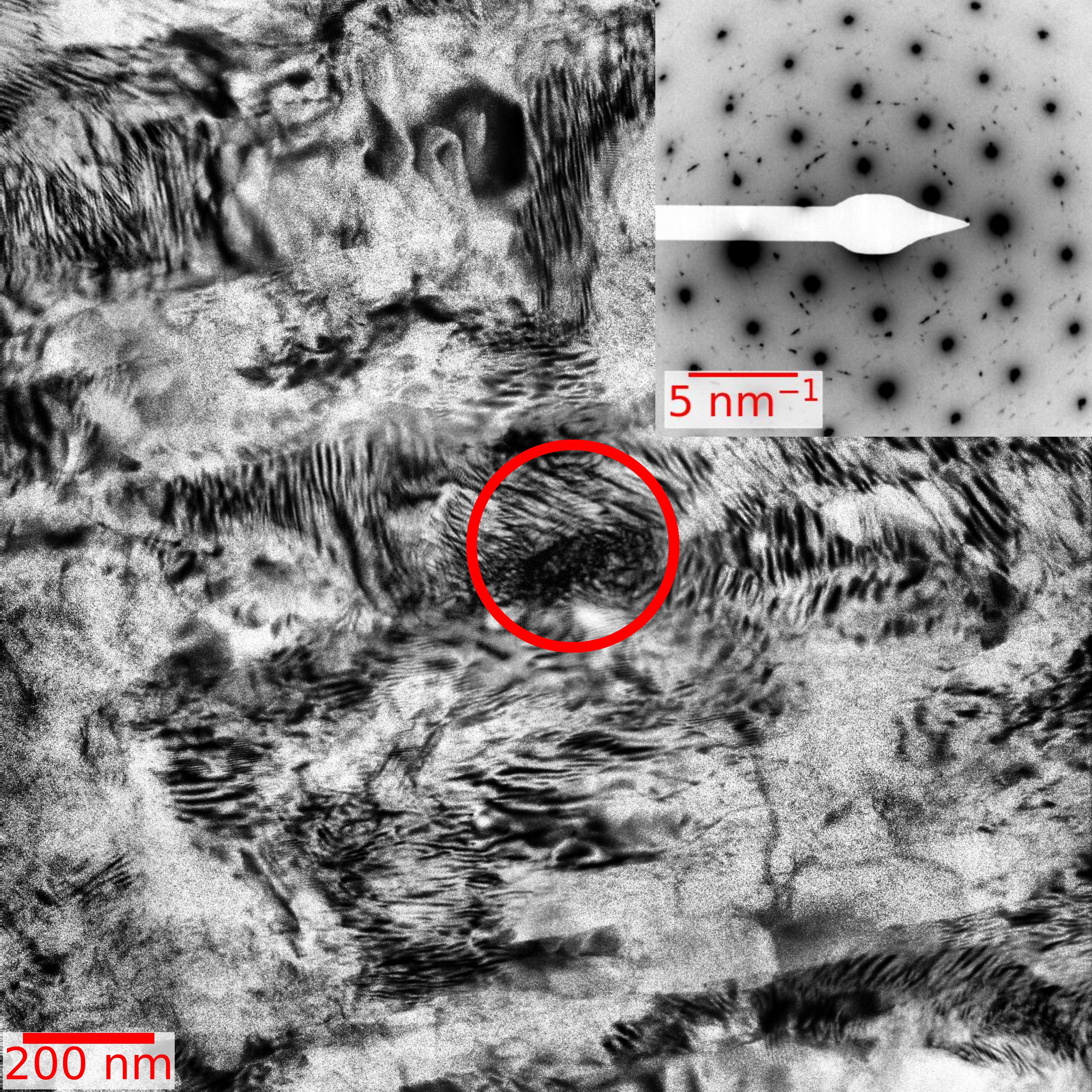}}
	\hfill
	\subfloat[\label{fig:20230531-mgmn-0rotn-a3-tem-twin} ]{\includegraphics[width=0.32\linewidth]{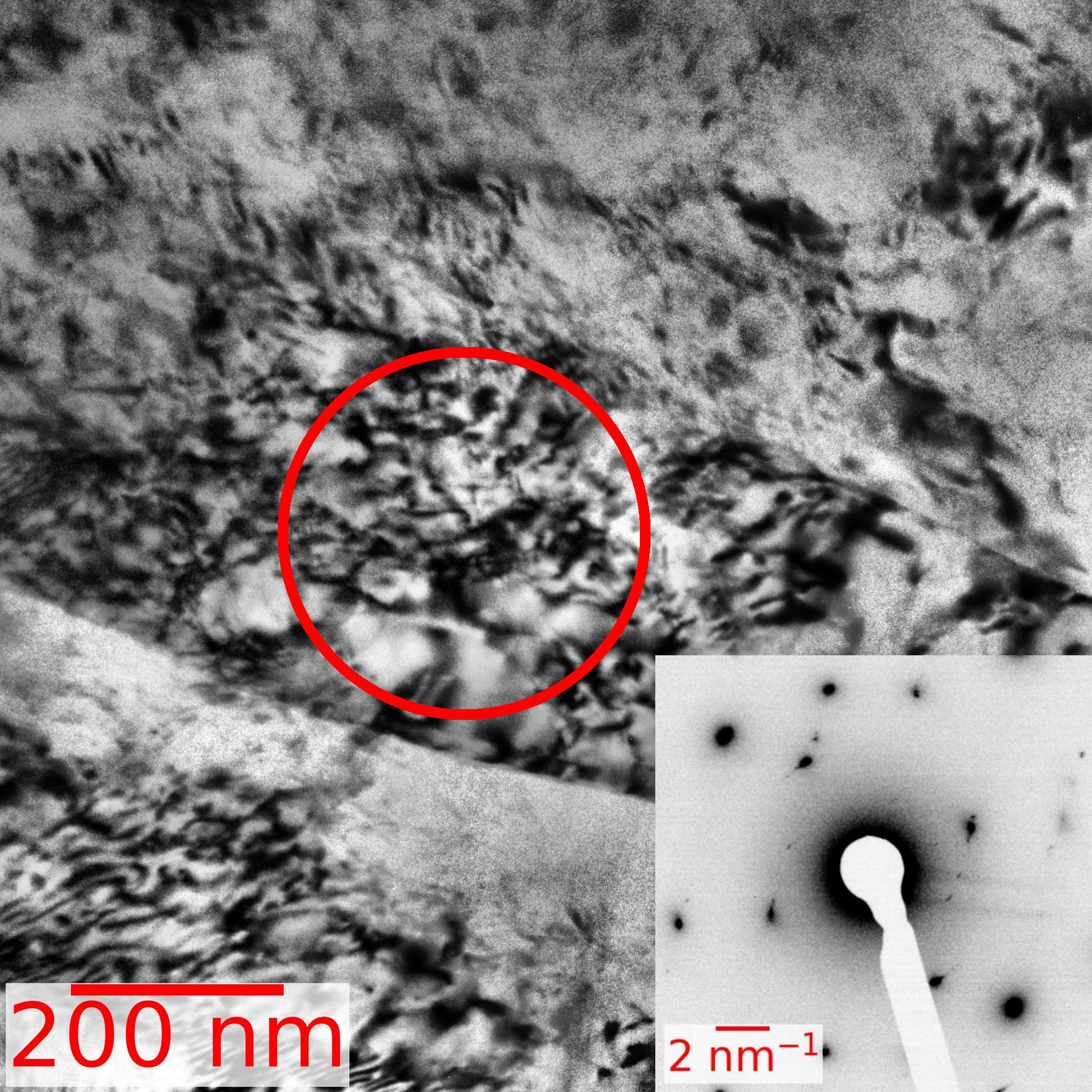}}
	\hfill\ 
	\subfloat[\label{fig:20230531-mgmn-0rotn-a3-tem-ufg}]{\includegraphics[width=0.32\linewidth]{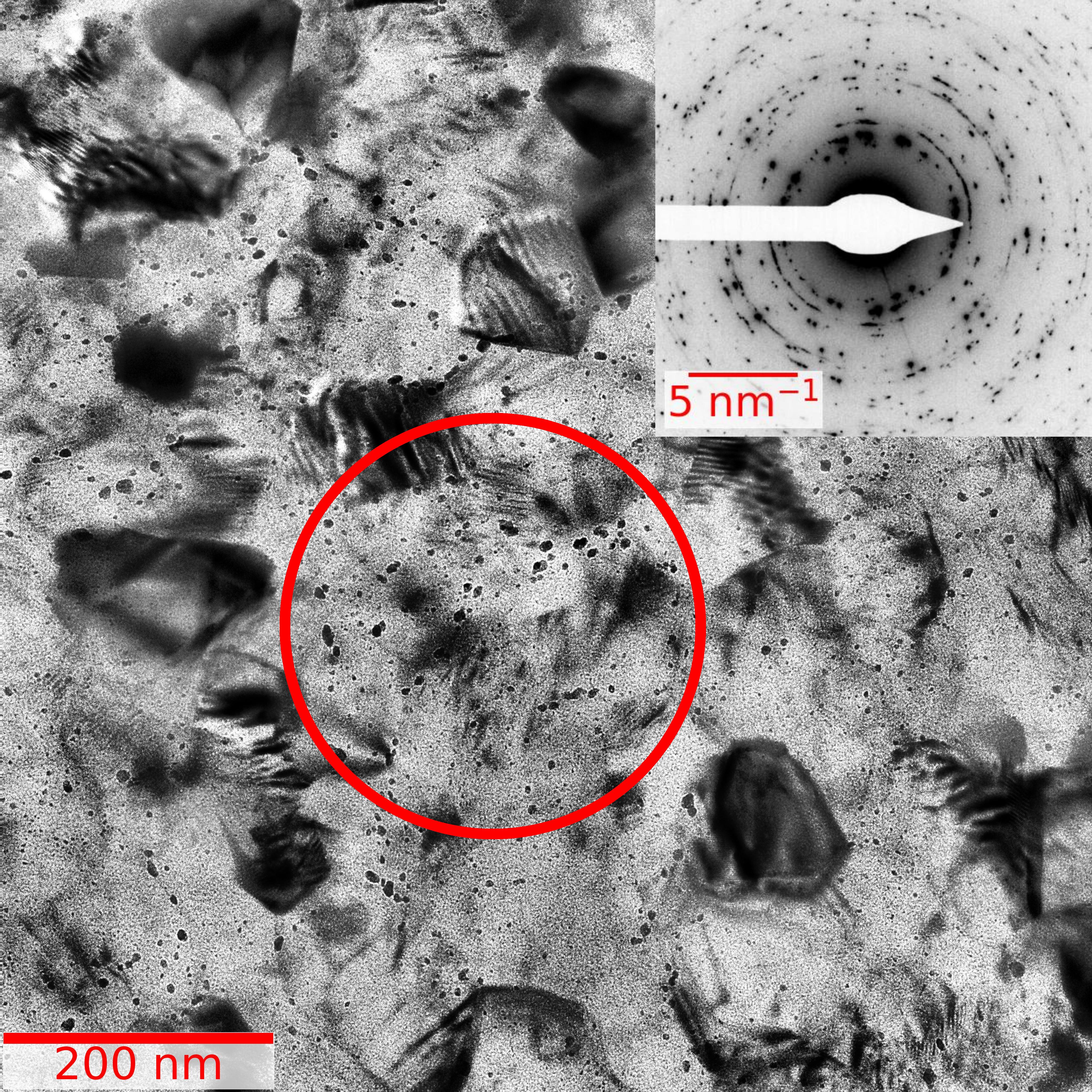}}
	\hfill\	
	
	\caption{Microstructures present in the as-compressed sample. The figure shows \gls{bf}-\gls{tem} images of the (a)  matrix, (b) twin and (c) ultrafine-grained regions, with the corresponding \glspl{sadp} inset in the image. While MgO reflections were often observed (and are visible in  (a)  and (b)),  the Mn 411 and 330  reflections appear only in the \gls{ufg} regions such as  (c). 
	 \label{as-received-tem}}
\end{figure*}

The microstructure of samples solution treated, quenched and compressed between the \gls{hpt} anvils was far more heterogeneous. Some regions contained  heavily-deformed, sub-micron grains of magnesium, as shown in Figure~\ref{fig:20230531-mgmn-0rotn-a3-tem-bulk}. The circled region indicates the location of the selected area aperture for the corresponding diffraction pattern (Fig.\ref{fig:20230531-mgmn-0rotn-a3-tem-bulk}) which is indexed to the 0001 zone. 
Other regions were dominated by broad twins as shown in Figure~\ref{fig:20230531-mgmn-0rotn-a3-tem-twin} and the accompanying \gls{sadp} in Fig~\ref{fig:20230531-mgmn-0rotn-a3-tem-twin}. The breadth of the twin (approximately 600\,nm) in this instance is consistent with  $11\overline{2}0$ ( ``$c$-axis tension'' ) twinning, which can occur readily at low stresses. A high density of dislocations is apparent within the twin and the surrounding matrix. 

 Ultrafine-grained regions were also common in this condition were, as can be seen in Figure~\ref{fig:20230531-mgmn-0rotn-a3-tem-ufg}. Close inspection of the circled region reveals the presence of a series of darker particles. The \gls{sadp} (inset)  obtained in this circled region has intensity at reciprocal distances matching not only magnesium but also with the 411 and 330 reflections of $\alpha$-manganese  both of which appear at a 2.03\AA. Given the complexity of the ring pattern, a more detailed analysis would be required to completely rule out the MgO (200) reflection (at 2.13\AA), however, it is notable that these additional reflections were not present in sub-micron or twinned regions
(Fig~\ref{fig:mgmn-tem-summary-sem}-\ref{fig:20230531-mgmn-0rotn-a3-tem-twin}), nor were they observed in the as-received condition (Fig~\ref{fig:mgmn-tem-summary}), only where the material had undergone more extensive grain refinement  (Fig~\ref{fig:20230531-mgmn-0rotn-a3-tem-ufg}).

\subsubsection{\Gls{hpt} deformed}

\gls{tem} and \gls{stem} observations confirmed that the \gls{hpt}-deformed Mg-Mn developed a sub-micron grain size 
 with mainly high-angle grain boundaries during \gls{hpt} deformation and retained this structure even at extreme strains. 
  The other notable feature was the presence of nm-scale particles concentrated along the grain boundaries. These particles will be shown to consist of Mn, and to play a vital role in stabilising the microstructure during further deformation.

\begin{figure}
	\centering
	\hfill\ 	
	\subfloat[\label{fig:bf-tem-1}]{\includegraphics[width=0.32\textwidth]{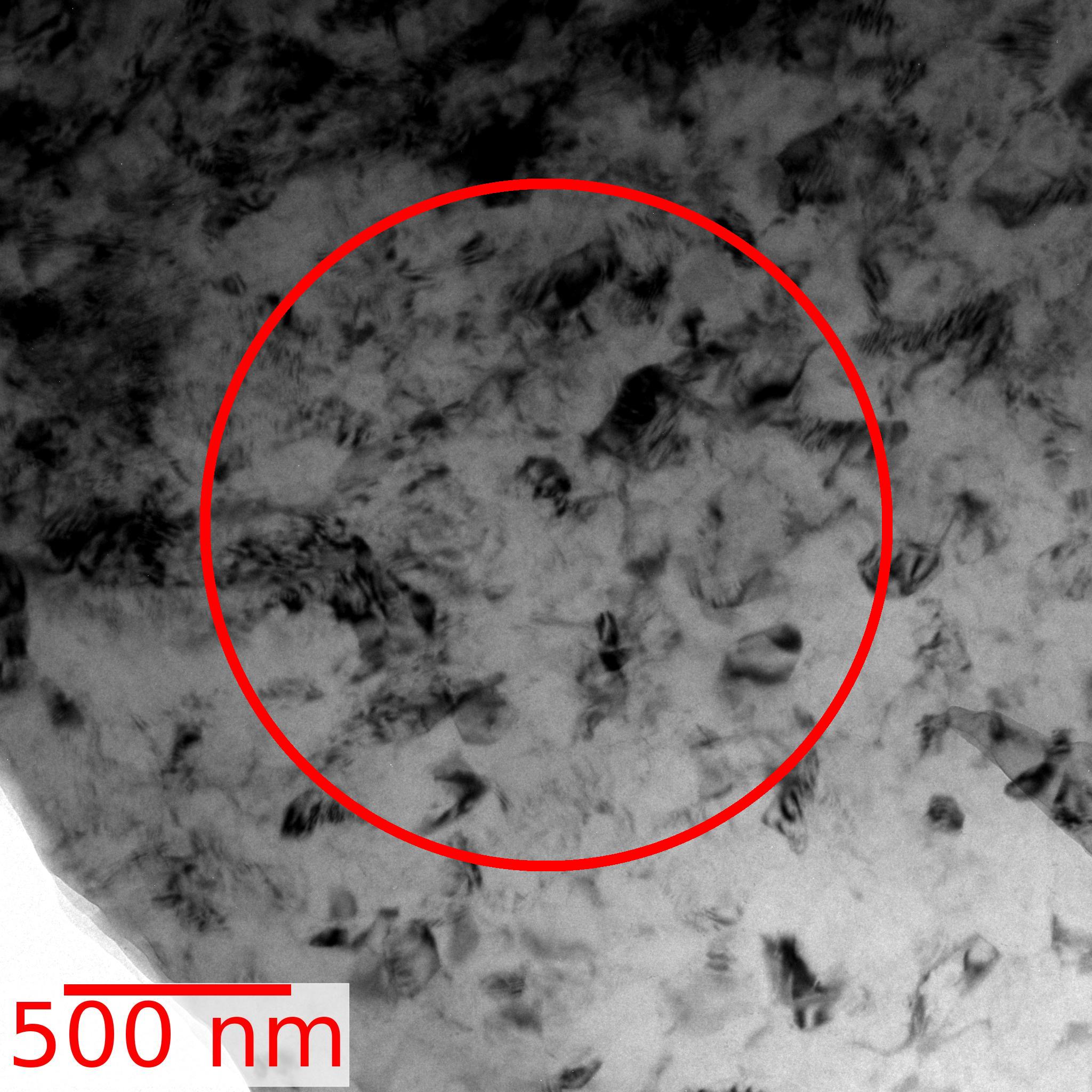}}
	\hfill
	\subfloat[\label{fig:bf-tem-sadp}]{\includegraphics[width=0.32\textwidth]{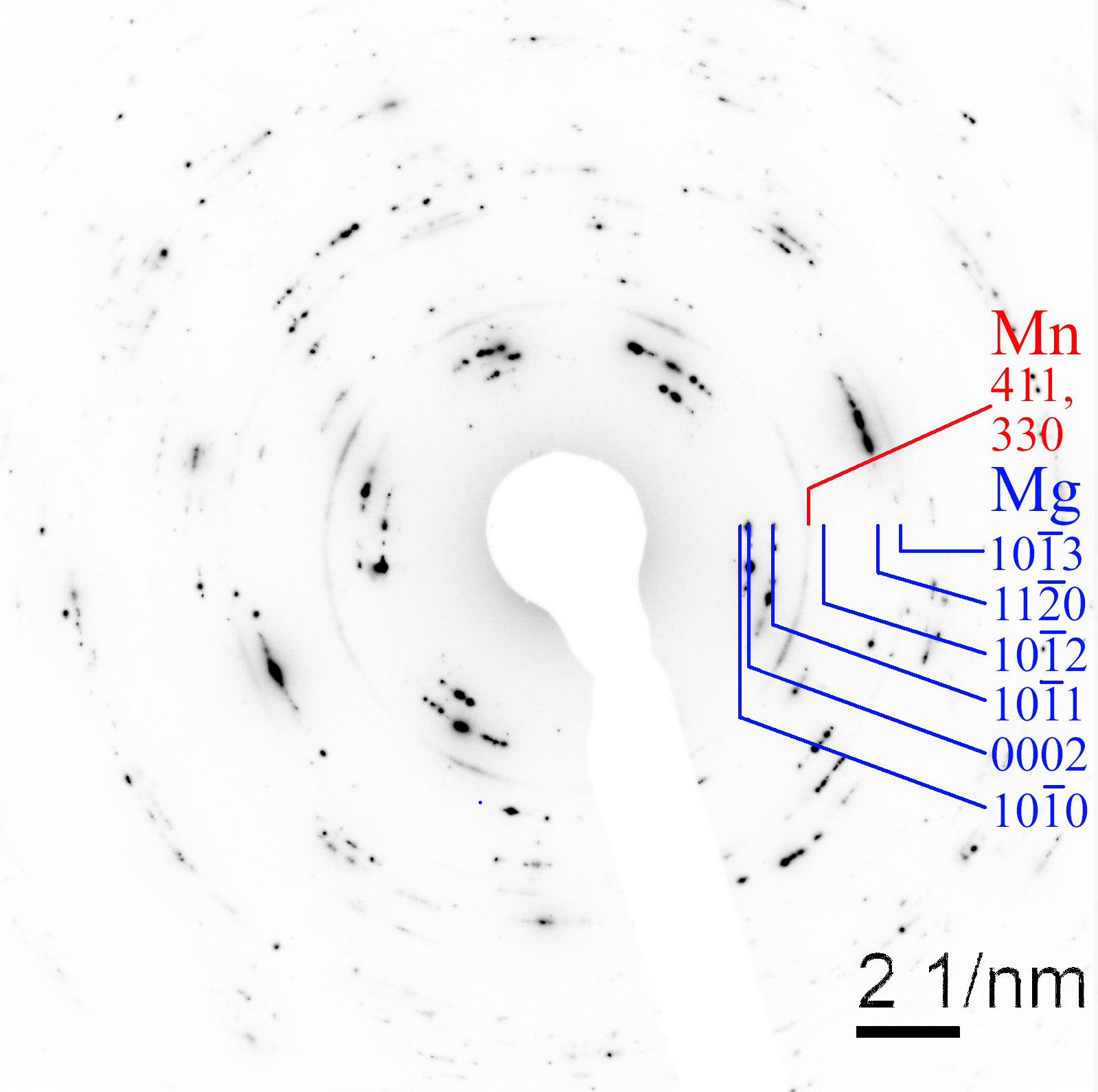}}
	\hfill\  	
	\caption{(a) \gls{bf} \gls{tem} image showing the microstructure of M1 after 1 rotation of \gls{hpt} deformation. 
		(b) The \gls{sadp} obtained with the aperture centred over the area circled in (a). In addition to Mg reflections the diffractogram shows a weak reflection attributable to the Mn (411) and (330) planes.	}
\end{figure}

\Gls{ufg} regions of the compressed sample had a surprisingly refined mean grain diameter of 110\,nm, however this does not take into account the inhomogeneous microstructure and the presence of remnant coarse-grained material. Further deformation resulted in the formation of an equiaxed ultrafine grained structure, as shown in Figure~\ref{fig:bf-tem-1} for an alloy deformed by one complete rotation. A selected-area diffraction pattern \ref{fig:bf-tem-sadp}, obtained with the aperture positioned over the red, circled area in \ref{fig:bf-tem-1}, showed reflections of a large number of Mg grains, with no obvious texture. In addition, there was a weak reflection with a spacing matching that of the Mn 411 and 330 planes. Given that the starting material was a bulk sample, and not a powder, and that \Gls{hpt} deformation tool place within a fully-sealed environment, the presence of oxides was ascribed to minor surface oxidation, not unexpected for a magnesium alloy. Aside these weak oxide peaks, no evidence for other phases was detected. 

\begin{figure*}
	\centering
	\hfill\ 
	\subfloat[0.5 rotations\label{fig-bf-stem-0p5}  ]{\includegraphics[width=0.32\textwidth]{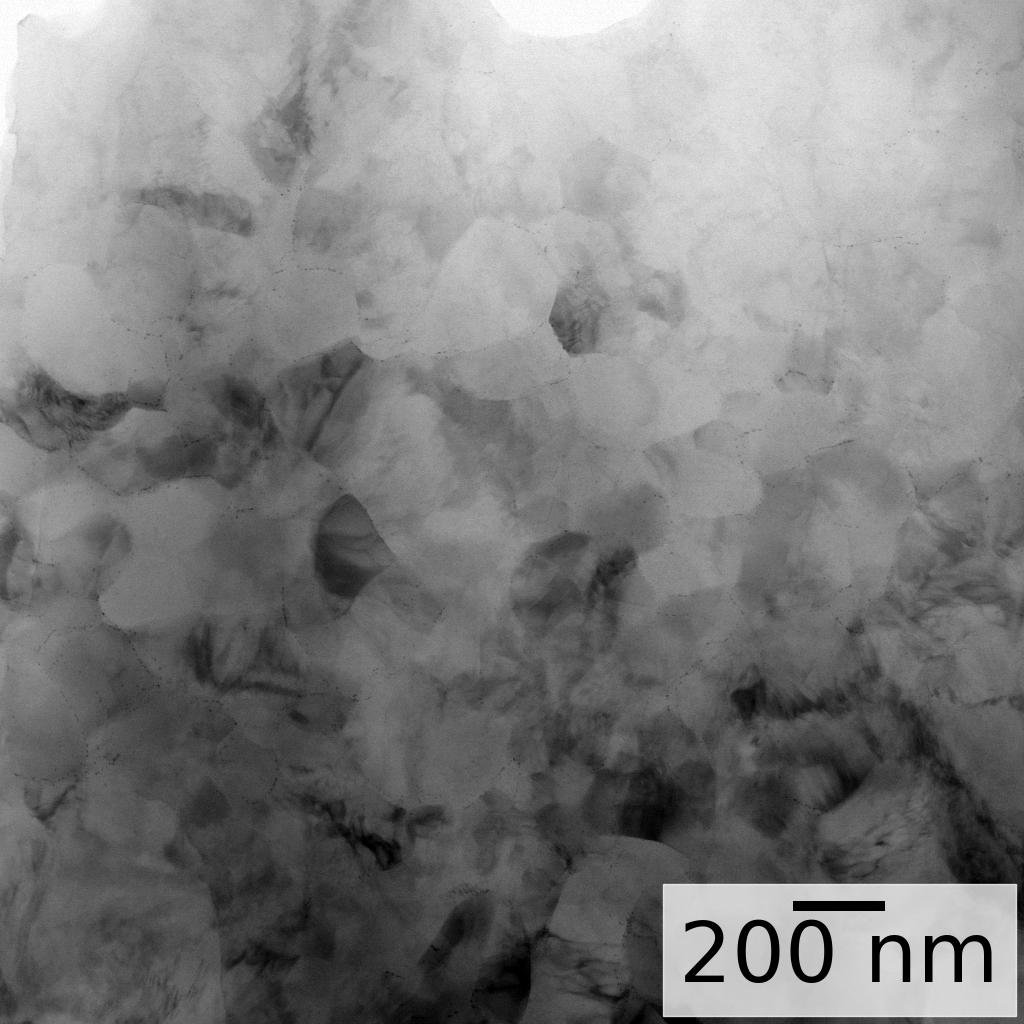}}
	\hfill\ 	
	\subfloat[5 rotations \label{fig-bf-stem-5} ]{\includegraphics[width=0.32\textwidth]{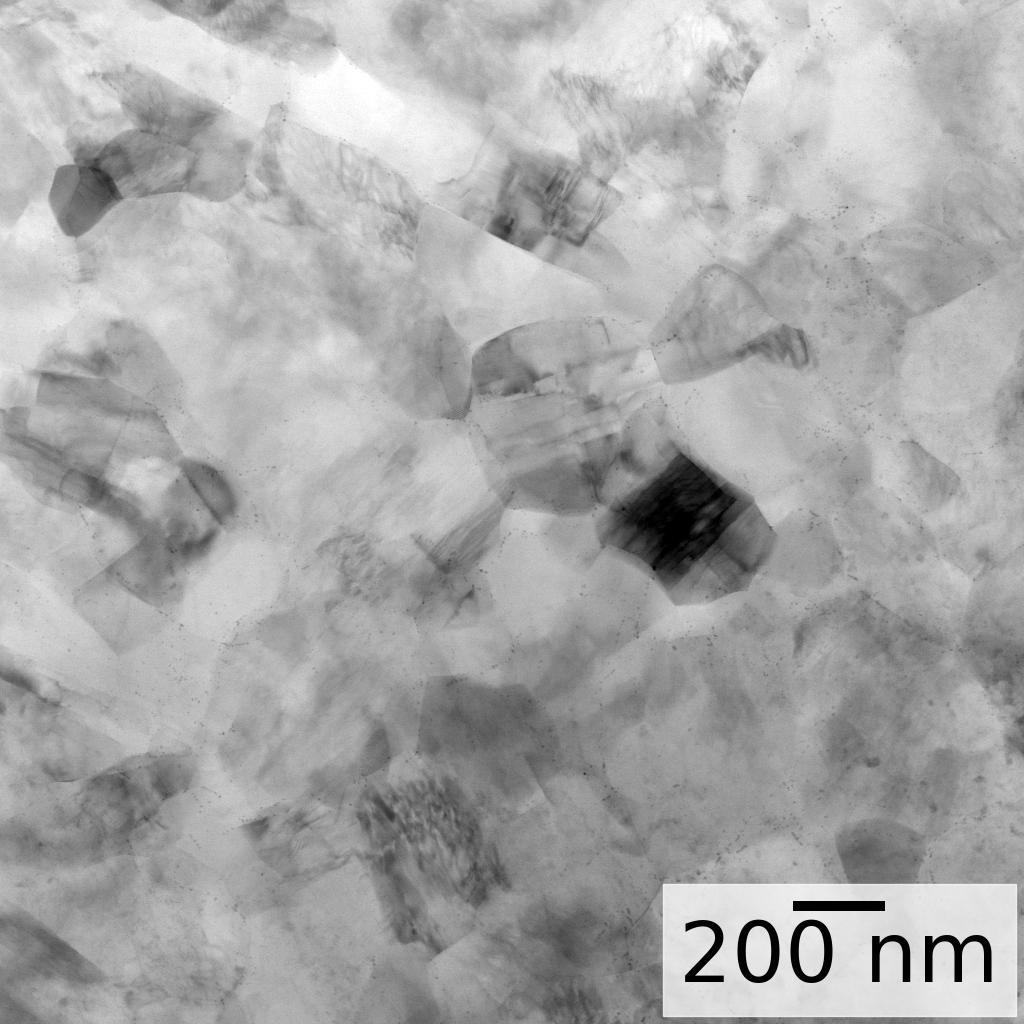}}
	\hfill\ 
	\subfloat[Grain size \label{fig-grain-size}]{\includegraphics[width=0.32\textwidth]{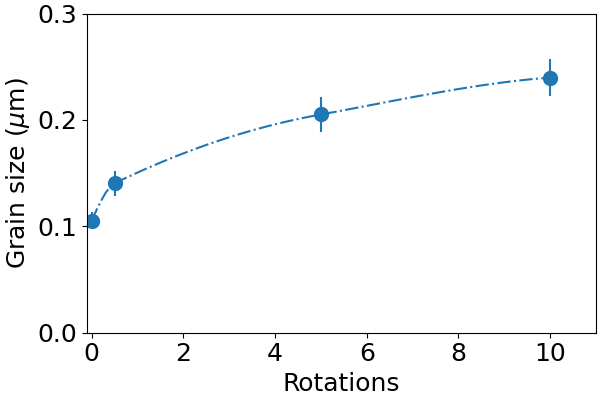}}
	\hfill\ 
	
	\caption{\label{fig-stem-hpt} \Gls{bf}-\gls{stem} images showing the grain structure of the Mg-Mn alloy, for (a)  0.5 rotations and (b) 5 rotations of \gls{hpt} deformation. The mean grain size as a function of the amount of deformation in shown in (c).} 
\end{figure*}

The grain diameter for each condition is plotted in Figure~\ref{fig-grain-size}, showing a minimum of 110\,nm for the \gls{ufg} regions formed after compression, and a value of 140\,nm for 0.5 rotations (at which point the entire microstructure was in an ultrafine-grained condition). This \gls{ufg} structure was retained even after extended \gls{hpt} deformation, as can be seen in Figure~\ref{fig-stem-hpt} which presents a series of \gls{bf}-\gls{stem} images for deformations of 0.5 and 5 rotations. The grain size increased gradually with further deformation, reaching 250\,nm after 10 rotations.
Despite the high strain to which the alloy was subjected, and the widespread strain contrast within the grains in Figures~\ref{fig-stem-hpt}, dislocations were not commonly observed.

\begin{figure}[hbpt]
	\begin{center}
		\hfill
		\subfloat[\label{fig-1rotn-tem-1}]{\includegraphics[width=0.32\textwidth]{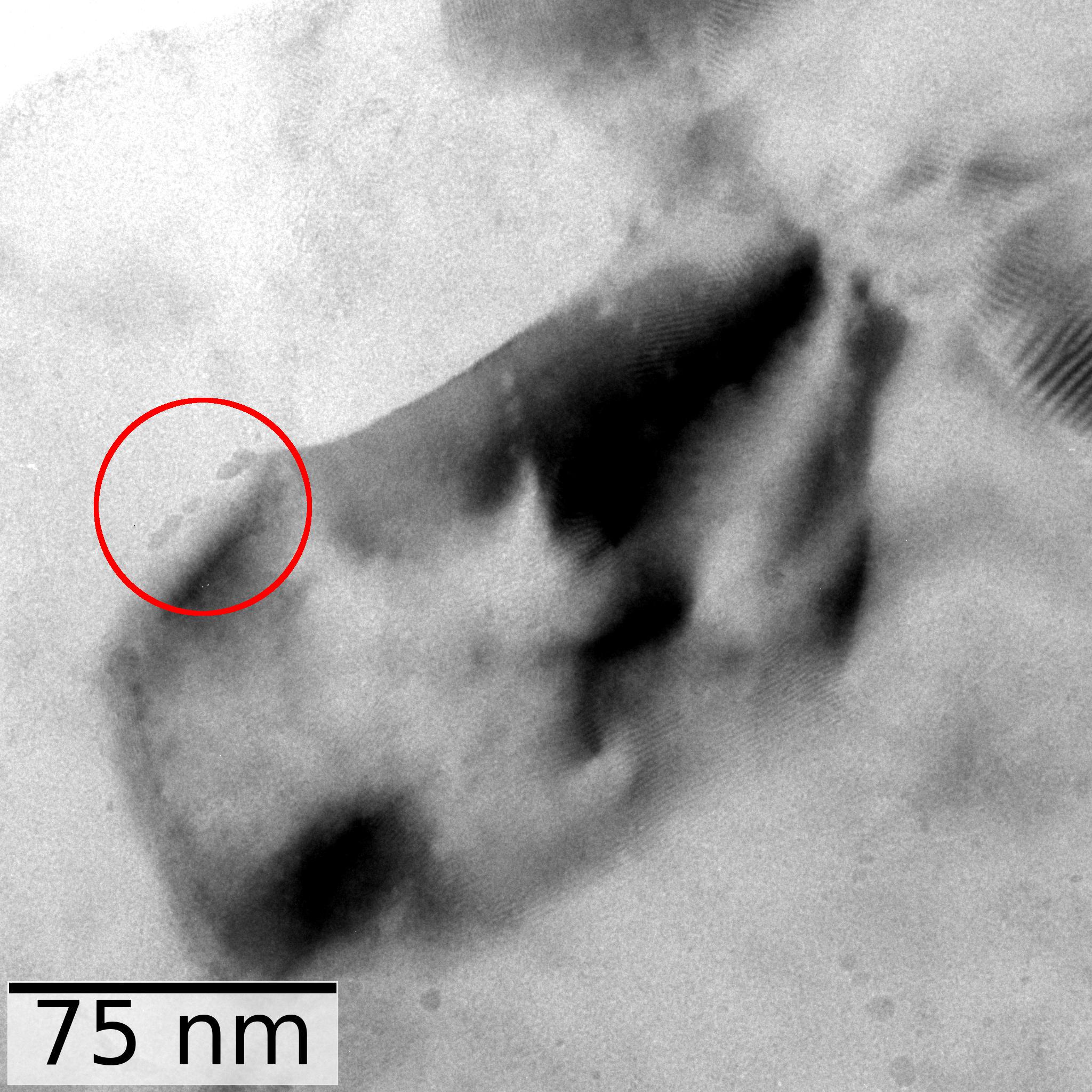}}
		\hfill
		\subfloat[\label{fig-1rotn-hrtem}]{\includegraphics[width=0.32\textwidth]{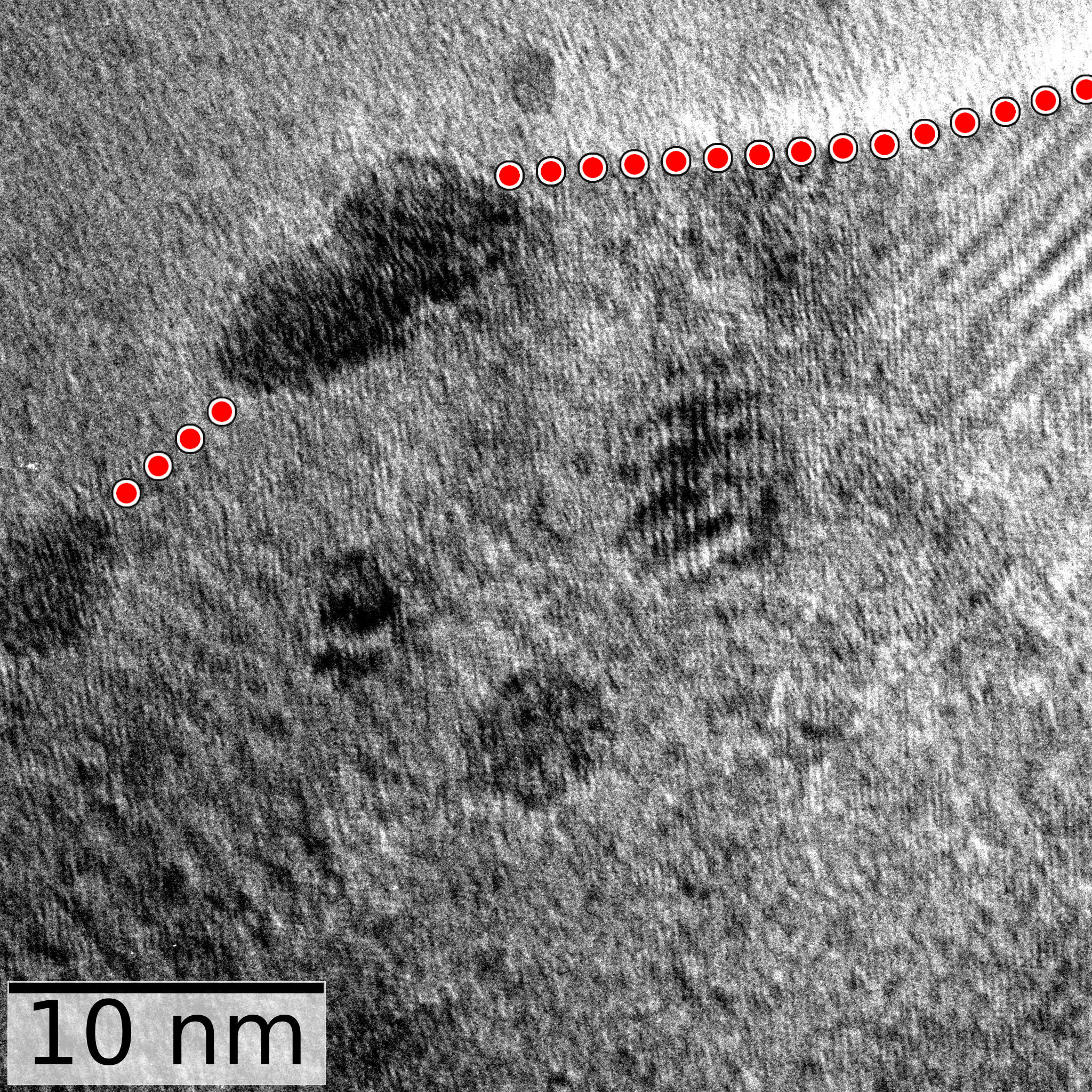}}
		\hfill\ 
		
\caption{\label{fig-tem-particles} Grain boundary particles in an M1 alloy deformed by 1 rotation of \gls{hpt}.
	(a)  \gls{tem} image showing grain boundary Mn precipitates surrounding a single Mg grain 
	(b) \gls{hrtem} image of Mn particles in the circled region of (a). The grain boundary is indicated by the dotted line.}
		\end{center}
	\end{figure}

\begin{figure}[hbpt]
	\begin{center}
		\hfill
		\subfloat[1 rotation \label{fig-1rotn-stem}]{\includegraphics[width=0.32\linewidth]{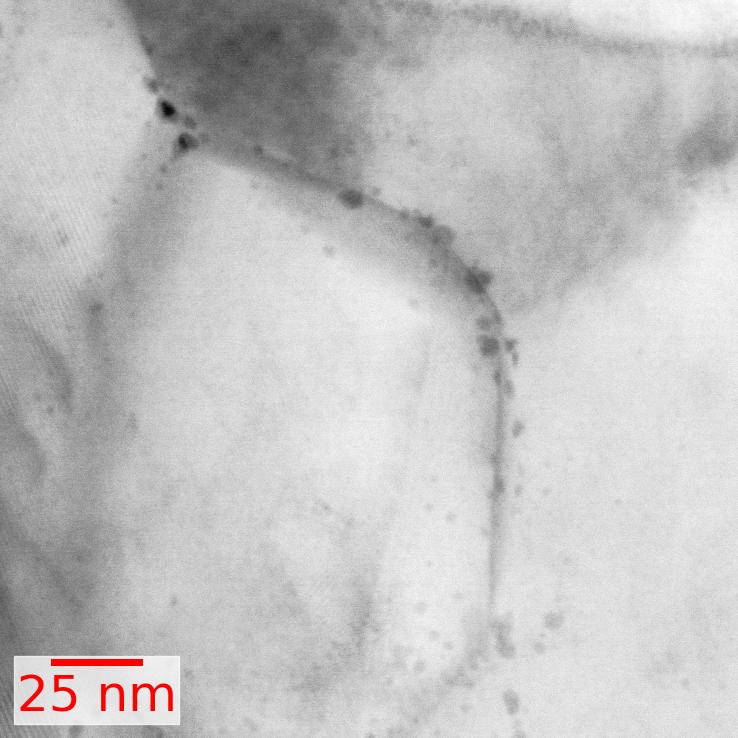}}
		\hfill
		\subfloat[10 rotations \label{fig-10rotn-stem}]{\includegraphics[width=0.32\textwidth]{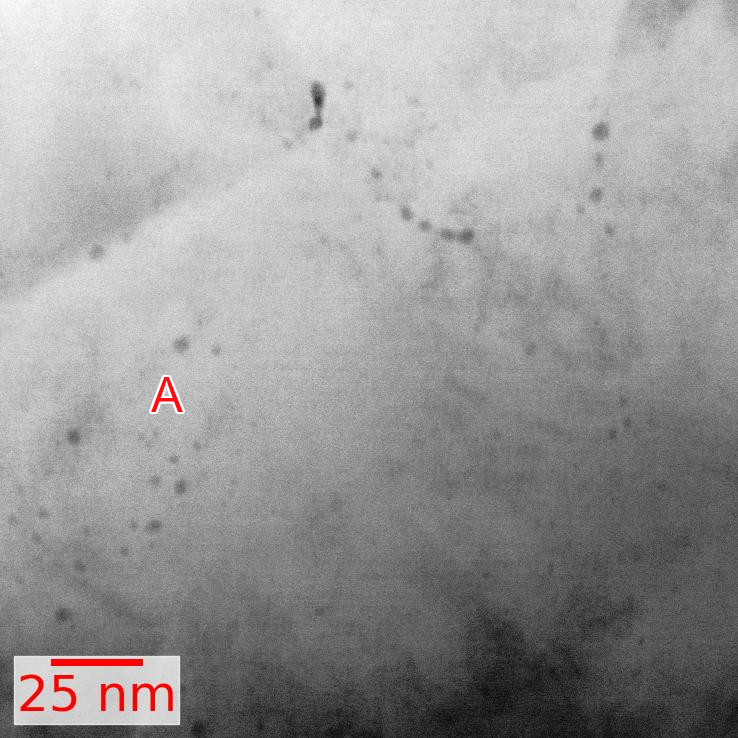}}
		\hfill\ 
		\caption{De-pinning of particles during extensive deformation.\label{fig-stem-depinning}
		(a)  \gls{bf}-\gls{stem} image showing Mn precipitates along  grain boundaries after 1 rotation of \gls{hpt}. 
		(b) \gls{bf}-\gls{stem} image in a sample deformed by 10 rotations of \gls{hpt}. Note the cluster of Mn particles at ``A" with no apparent association with a grain boundary.
 }
	\end{center}
\end{figure}

Higher magnification \gls{tem} imaging showed that the particles were primarily  arranged along the grain boundaries. Figure~\ref{fig-1rotn-tem-1} shows a micrograph of a Mg grain in a sample deformed by 1 rotation of \gls{hpt}. Circled regions indicate the locations of 2-5\,nm particles along and adjacent to the grain boundaries. The high resolution image
(Fig.~\ref{fig-1rotn-hrtem}) 
shows one of the circled regions in more detail, making it clear that a number of the particles lie precisely along the grain boundary, which is indicated by a dotted line. Measurements from multiple micrographs indicated a high degree of variability in the particle size with a mean diameter of approximately~2.5\,nm. 

After prolonged deformation, particles were also observed within the intragranular region. This can be seen in the \gls{bf}-\gls{stem} images in Figure~\ref{fig-stem-depinning} which compares the microstructure after (a) 1 and (b) 10 rotations. In the former case the Mn particles are found almost exclusively along recognisable grain boundaries, however in the latter, a number of particles (indicated by the letter ``A" in the figure) show no clear association with any boundary. Significantly, the arrangement of these particles suggest they may have formed on a grain boundary which has since migrated, consistent with the observation that the grain sizes increase during deformation.

\section{Discussion}

\subsection{Grain refinement}

It is useful to examine the response of the MgMn alloy relative to systems reported in the literature which share common features. Given the low alloy content (0.63 at.\%), the as-solutionised condition  is worthwhile to compare with pure magnesium. The process of grain refinement in the MgMn alloy was initially similar to that in pure magnesium \cite{Silva2019,QiaoZhao2014,Edalati2011}, with the as-compressed material developing an inhomogeneous microstructure containing both coarse and ultrafine grains and  rich in tensile twins \cite{QiaoZhao2014}. Like pure Mg, the MgMn alloy also underwent a rapid increase in hardness. The alloy reached a maximum of 88\,\gls{hv} for equivalent strains of around 10 (corresponding to  $\sim$0.5 rotations in the present work) with no further hardening observed. 

Distinct differences, however, became apparent in the behaviour at higher strains, once precipitation of $\alpha$-Mn progresses. In pure magnesium, a steady state was reached, with a constant hardness\cite{Silva2019,QiaoZhao2014,Edalati2011} and consistent texture \cite{Bonarski2008}.
In contrast the manganese-containing alloy 	underwent a gradual and continuing increase in grain size (see Fig~\ref{fig-grain-size})   coupled with a decrease in hardness (see Fig.~\ref{fig-hardness}).
Pure Mg also developed a bimodal grain structure distribution, with some grains up to $\sim1.5\mu$m in diameter\cite{Edalati2011,Bonarski2008}, a feature which was notably absent in the present system. This observation which points to differences in the dynamic recrystallisation and growth, presumably due to the presence of the Mn precipitates.

Once the precipitation of Mn has commenced, comparisons can also be drawn with \gls{hpt} of two-phase systems. The Mg-Mn system is immiscible at room temperature 
and the $\alpha$-Mn particles are present as isolated, hard particles within a continuous, softer matrix. 
The microstructural evolution of similar systems during \gls{hpt} has been  well-studied and where the volume fraction of the hard phase is low, as in this case, the deformation is primarily localised in the soft phase, resulting in grain refinement of this phase. The evolution of such systems typically involves a three-stage sigmoidal hardness increase, concluding in a steady state in which the hardness reaches a plateau after around 8-10 rotations of \gls{hpt} deformation \cite{KormoutPippan2016}. This behaviour is distinctly different from that of the MgMn alloy.

Parallels can also be drawn with systems where precipitation occurs during deformation.
	Unfortunately, studies of dynamic precipitation during \gls{hpt} of other alloy systems are rare, with Al-Cu being a notable exception. In room temperature of solutionised \gls{hpt} Al-3 wt.\%Cu the hardness reached a steady state after 3.5 revolutions \cite{Hohenwarter2014}, which is consistent with other reports \cite{Nasedkina2017} in which the grain size reached a steady state, with no measurable change observed between equivalent strains ranging from 16 to 310. So although the present system applies room-temperature deformation to a supersaturated solid solution, which undergoes grain boundary precipitation, as is the case with Al-Cu, the microstructural and hardness evolution is distinctly different.

Two questions therefore arise regarding the microstructural evolution of the Mg-Mn alloy during \gls{hpt}, firstly, why does it initially develop such fine grain sizes, and secondly why do the grains subsequently coarsen? The absence of a bimodal grain structure is also of interest. The reasons underlying this atypical behaviour are explored in this section.

Firstly, the extent of grain refinement in the early stage of \gls{hpt} deformation demonstrates that Mn can provide an exceptionally high resistance to grain growth. This is best illustrated by comparing the minimum grain sizes achieved by \gls{spd} of Mg and its alloys. As a baseline, pure Mg is limited to grain sizes in the range of 0.6--1.0$\mu$m\cite{Edalati2011,Sulkowski2020,QiaoZhao2014}. Finer grain sizes can be achieved through deformation of alloys, and higher solute content generally results in greater resistance to coarsening, however, the minimum grain size of 140\,nm in this work is comparable with results  for \gls{hpt} and \gls{ecap} of AZ \cite{Silva2019, XiaWang2005} and ZK alloys. It should be noted that the values reported in this work are actual grain sizes, rather than (typically lower) coherent domain sizes as would be measured by \gls{xrd} or \gls{df}-\gls{tem}.
As can be seen in  Table~\ref{tab-volume-fraction},  the solute content for M1 is considerably lower and the maximum volume fraction of precipitates is almost an order of magnitude lower than for AZ 31.

\begin{table}
	\centering
	\caption{Precipitate volume fraction and minimum reported grain size for selected magnesium alloys. Volume fraction values are from experimentally-obtained data under conditions conducive to extensive precipitation.\label{tab-volume-fraction}}
	\begin{tabular}{lp{9ex}p{9ex}p{10ex}}
		\toprule
		Alloy &   solute &   $v_F$   & d                      \\
		& (at.\%)  &  (\%) & ($\mu$m)   \\
		\midrule 
		CP-Mg  &  ---			&---  & 0.206\cite{Silva2019}	\\		
		M1    &   0.63          & 0.32 & 0.1*\\ 
		AZ31  &     3-4       & $\sim$0\cite{ZhangZha2021} & 0.119\cite{Silva2019}, 1\cite{XiaWang2005}\\
		AZ61  &     6-7        & 15\cite{Mohseni2021} & 0.22\cite{Harai2008}\\
		AZ91  &     9-10        & $\sim$15\cite{Zeng2013}- 23\cite{ZhangZha2021} & 0.141\cite{Silva2019}\\
		ZK60  &     6-7        & 25\cite{WangYang2022} & 0.133 \cite{Silva2019} \\
		\bottomrule
	\end{tabular}
	
	*Present work
\end{table}

Secondly, it is evident that the present system compares favourably with previous reports of grain refinement in other Mg-Mn alloys, even those with significantly higher alloying content. For example, extrusion of an M3 alloy at 250$^\circ$C yielded an average grain size of 1.5$\mu$m\cite{YuTang2018}, substantially greater than reported here, despite the much higher particle volume fraction. This is likely to derive from the high temperatures employed during extrusion permitting more rapid grain growth than room-temperature \gls{hpt} as used in this work.
\Gls{spd} via \Gls{ecap} was more effective, producing grain sizes of 0.5\,$\mu$m in 1.5 wt\% Mn \cite{SvecDuchon2012}, but still well above the grain refinement achieved by \gls{hpt}.

The fundamental difference in this work is that the starting material is in a solutionised state, with a negligible volume  fraction of pre-existing Mn particles. Any pinning particles must therefore be formed \textit{in-situ}, during deformation. The \gls{tem} observations of numerous grain boundary precipitates (See Fig~\ref{fig-tem-particles},\ref{fig-stem-depinning}) demonstrate that extremely fine (2-5\,nm) particles are indeed able to nucleate during deformation, and that a significant fraction of these remain at the grain boundaries, even after extensive deformation. The resulting microstructure resembles that reported by Yu et al. \cite{YuTang2018}, but with the Mn particles here being finer and also very heavily concentrated at the grain boundaries. 

This suggests that solute manganese plays a similar roll to its use in creep-resistant Mg alloys, where particles are required to dynamically precipitate on, and pin, mobile dislocations\cite{CelikinKaya2012, CelikinKaya2012a}.  Remarkably, despite the extreme strain experienced by the alloy, and the well-known acceleration of diffusional processes during \gls{hpt} there was no obvious change in the particle size during deformation.

The occurrence of precipitation within mere minutes of deformation is understandable given the  well-established observation that diffusion and diffusional processes are accelerated by orders of magnitude during \gls{spd} \cite{StraumalBaretzky2004,StraumalMazilkin2012,SauvageWetscher2005}. This can naturally result in accelerated precipitation, as for example in solution-treated Al-Cu alloys where Cu-rich phases 
formed during deformation to comparable strain values during room-temperature \gls{hpt} \cite{Hohenwarter2014,Nasedkina2017}.

The reason for subsequent grain growth during deformation is less obvious. Deformation temperature has a strong influence on grain refinement, however frictional heating during \gls{hpt} appears unlikely to be a significant factor here. Finite element models \cite{Figueiredo2012} predict very modest temperatures of up to $\sim55^\circ C$  during \gls{hpt} deformation of softer metals such Al, and by extension Mg. Similarly, experimental studies measured temperature increases of at most 15$^\circ$C for Al deformed under similar condition, and it was noted that for all metals  tested, the temperature increase was minor in comparison with the corresponding melting temperature \cite{EdalatiMiresmaeili2011}.
It is therefore necessary to look to factors other than frictional heating to explain the grain growth during extension \gls{hpt} deformation.

If the particles are considered as classical Zener pinning sites, the particles are assumed to be homogeneously distributed and their effectiveness is determined only  by the particle size and volume fraction. The limiting grain size ($D_Z$), i.e. where the pinning pressure is sufficient to halt grain growth can be calculated from the equation: \cite{Robson2011}
\begin{equation}
D_Z =  \frac{4r}{3V_f}   
\end{equation}
which for uniformly distributed particles of radius, $r$, present with volume fraction of $V_f$, yields a critical grain size of 8$\mu$m. This would not explain the fine grain size achieved during the initial stages of deformation, even assuming that all available solute was partitioned into particles.

More detailed models of particle pinning \cite{ManoharFerry1998,WoldChambers1968} have taken into account the preferential formation of particles along grain boundaries, as is the case here, expressing this is a multiplicative factor, $k$, for the particle pinning pressure, with
\begin{equation}
D_Z(k) =  \frac{4r}{3kV_f}   
\end{equation}
for incoherent particles.  This reflects the greater number of interaction between such particles and the boundaries, compared to the classical assumption of a homogeneous distribution.

Dynamic precipitation of Mn particles  would result in a concentration of particles along grain boundaries as in Figures~\ref{fig-tem-particles} and \ref{fig-1rotn-stem}, which provides strong  resistance to grain growth, expressed as a high $k$ factor. However, under the extreme conditions of \gls{hpt} deformation, the pinning pressure can be overcome, and once de-pinning occurs (Fig~\ref{fig-10rotn-stem})  the resistance to grain boundary migration will be reduced, until such time as further precipitation occurs along the migrated grain boundary. Over time, the driving force for precipitation will also decrease as the level of supersaturation decreases, meaning that dynamic precipitation will become less effective as deformation proceeds. The  initially high, but gradually decreasing, $k$ value would explain the more extensive initial grain refinement for the solutionised Mg-Mn; due to the strong localisation of particles on the grain boundaries,  compared to alloys in which significant precipitation has already occurred \cite{SvecDuchon2012,YuTang2018}  and ii) secondly the increasing grain size as deformation proceeds.

\subsection{Strengthening mechanism}

\subsubsection{Hall--Petch strengthening}
\begin{figure*}
	\centering
	\includegraphics{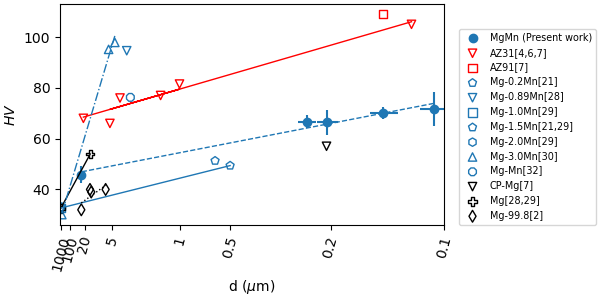}
	\caption{Hall--Petch plot of hardness vs reciprocal grain diameter for  magnesium alloys. The errors bars for the present work indicate the standard deviation of the hardness value.}
	\label{fig:hallpetchmgmn}
\end{figure*}

The hardness of the alloy followed a Hall--Petch relationship, as is typical for severely plastically deformed Mg alloys \cite{QiaoZhao2014}. 
Figure~\ref{fig:hallpetchmgmn} shows microhardness values plotted against grain size, $d$ for the present alloy and a number of magnesium alloys reported in the literature. 
Where hardness data was not available, values were estimated from the published yield-strength values, using the equation $H = 0.38 (\sigma_y + 45) $ as proposed by C\'{a}ceres et. al \cite{CaceresGriffiths2005}.

The Hall--Petch gradient was similar to that of AZ31 and  to M-series alloys containing up to 1.5 wt.\% Mn \cite{YuTang2014,YuTang2019}. 
The hardness of Mg alloys with 2 wt.\% and 3 wt.\% Mn alloys (i.e.. close to or beyond the maximum solid solubility) had a higher Hall--Petch coefficient, showing a much stronger dependence on grain size  \cite{YuTang2018}. This is consistent with reports on the AZ series, where the hardness of alloy with extensive precipitation (AZ61\cite{Harai2008},AZ91\cite{Al-Zubaydi2016}) exceeded that of those with little or no precipitation(AZ31\cite{XiaWang2005,Silva2019}).

\nocite{Sulkowski2020, XiaWang2005, Harai2008, XuShirooyeh2013,Silva2019, SvecDuchon2012, YuTang2014, YuTang2018,YuTang2019, SomekawaBasha2018}

\subsubsection{Other strengthening mechanisms}
Strengthening mechanisms other other Hall--Petch strengthening are expected to play considerably less significant roles in the \gls{hpt} M1 alloy. 

Although dislocation strengthening is generally important in heavily-deformed pure metals, \gls{spd} of low-melting metals such as Mg is unlikely to adhere to this trend \cite{StarinkCheng2013}. An \gls{ecap} study on pure magnesium estimated dislocation hardening contributed only $\sim$15\,MPa to the strength for grain sizes around 200\,nm and only $\sim$6\,MPa for $\sim$100\,nm grains \cite{Balogh2010}. This is due to the high homologous temperature and subsequent dislocation annihilation in low-melting alloys, which can be extensive as to make accurate measurements of the dislocation density difficult. This is particularly the case in the \gls{ufg} regime, as both theoretical \cite{Kato2014} and experimental studies \cite{Conrad1967}	indicate that the dislocation density varies inversely with grain size.
Reductions in the dislocation density of Mg have been reported after 1 rotation of \gls{hpt} \cite{Edalati2011} and \gls{pals} measurements on room-temperature \gls{ecap} AZ31 alloy found a dislocation density of $1.6\times10^{13}$m$^{-2}$ \cite{Straska2014}. Taking this value as a guide, we would expect roughly 200\,nm line length, or only two dislocations, per 120\,nm grain. Indeed, the electron micrographs presented in this work show little  evidence of widespread dislocation activity.

The location of the particles makes it 
	  unlikely that dispersion strengthening played a significant role in the \gls{hpt}-deformed alloy. It has been established that Mn precipitation provides little reinforcement of the mechanical properties of coarse-grained Mg-Mn due to the coarseness of the distribution\cite{Borkar2012}. In this \gls{ufg} Mg-Mn, however, the precipitates are strongly concentrated  at grain boundaries, providing good pinning but little impedance to dislocation motion. Were this not the case, intragranular precipitation on this scale would potential provide potent strengthening:
Assuming full precipitation of Mn, and calculating the interparticle spacing, $L$, for spherical particles of radius, $r$ and volume fraction, $v_f$.
\begin{equation}
L = r \sqrt[3]{ \frac{4 \pi}{3 v_f} }
\end{equation}
gives a value of  25--30\,nm which should represent a substantial impediment to dislocation motion, something which is not evident in the hardness curves. Even after 10 rotations there is a strong concentration of particles along the grain boundaries and extensive particle-free regions within the intragranular regions. 

Lastly, although it was not possible to directly measure the solute content, there is good reason to expect that solid solution strengthening will be minimal.
Previous studies have found the effect of Mn to be an order of magnitude less than Al\cite{CaceresRovera2001} and Sn\cite{ShiChen2011} in magnesium, and two orders of magnitude less than Y, Gd\cite{GaoChen2009} and Zn\cite{YuTang2019}. This was born out in the observation that  Mn in solid solution did not limit coarsening during homogenisation\cite{YuTang2018}.

\subsubsection{Summary}

It is clear from the relatively low hardness in comparison to other \gls{spd} Mg alloys (Fig~\ref{fig:hallpetchmgmn}) that \gls{hpt} M1 has little potential in applications demanding high strength. Poorer than expected mechanical properties were also noted by Svec and Duchon in their \gls{ecap} alloy, where the \gls{uts} reached only 140\,MPa for 1.5 wt\% Mn with a grain size of 0.6$\mu$m \cite{SvecDuchon2012}. This was attributed to the formation of strong texture during \gls{ecap}. In contrast Yu \textit{et. al}\cite{YuTang2018} reported higher strength and good ductility in extruded samples, where there was sufficient Mn to weaken the basal texture.

However, \gls{ufg} magnesium is also of interest in non-structural roles such as in hydrogen storage and biomedical applications. Work is currently underway to evaluate the thermal stability and mechanical properties of this \gls{hpt}-deformed material for such applications. 

The optimum thermomechanical treatment to generate an \gls{ufg}  Mg-Mn alloy involves solution treatment followed by a brief exposure to severe plastic deformation; in the case of \gls{hpt} less than 0.5 rotations would appear sufficient. The resulting material should by characterised by moderately high hardness ($\sim$88$\pm$4\,\gls{hv}) with a grain size of approximately 150\,nm, and a sparse distribution of Mn particles along the grain boundaries.

It is worth noting that the required levels of deformation can be achieved via continuous high-pressure torsion (CHPT)  \cite{Edalati2010,Edalati2012}, which would open up the possibility for the production of continuous strands of \gls{ufg} magnesium wire. Such a material could have  potential in biomedical applications as a bioabsorbable structural support.

\section{Conclusions}

A \gls{ufg} magnesium alloy has been produced via \gls{hpt} of a solutionised binary magnesium-manganese alloy. Grain refinement was rapid, with ultrafine-grained regions (mean grain size 110\,nm) forming on initial loading and an overall \gls{ufg} microstructure with an average grain of 140\,nm becoming  established after 0.5\,rotations of \gls{hpt} deformation.
Subsequent deformation did not lead to a steady-state, but rather an 
atypical, gradual increase in grain size without the development of a bimodal grain structure. Grain-refinement and the retention of a fine, uniform grain size was attributed to the dynamic precipitate of nanometer-scale Mn particles which populated the grain boundaries and triple points and acted as pinning sites. 

\section*{CRediT authorship contribution statement}

\textbf{JMR:} Conceptualization, Formal Analysis, Investigation, Methodology, Software, Validation, 
Visualization, Writing – original draft, Writing – review \& editing
\textbf{AH:}
Investigation,
Resources, 
Writing – review \& editing,

\section*{Acknowledgments}
The authors would like to thank Dr. M. Alderman of Magnesium Elektron for providing the material used in this investigation. Ion polishing of the \gls{tem} samples was assisted by C. F\"orster, (Helmholtzentrum, Berlin). The \gls{sem} observations were conducted by P. Ocano-Su{\'a}rez (\gls{bam}). \gls{stem} measurements were carried out with the assistance of C. Prinz (\gls{bam}) and utilised the electron microscopy center at \gls{bam}. This work was funded solely by the Bundesanstalt f\"ur Materialforschung und -pr\"ufung (\gls{bam}). 

\section*{Data availability}
Data will be made available on reasonable request.


\end{document}